# New theoretical model of the dynamic anomalies in HTc superconductors

J. Sosnowski
SEP, Warsaw, Poland
sosnowski.jacek@wp.pl

*Abstract – New theoretical model of the dynamic current-voltage characteristics anomalies in HTc superconductors in slowly varying magnetic field has been proposed. The model is based on an analysis in details of the magnetic flux penetration into HTc superconducting slab in such case. The comparison of model with previous experimental data has been presented. Theoretical analysis resulting from a new solution of diffusion equation is given as well as other one based on phenomenological critical state model.*

## I. INTRODUCTION

HTc oxide superconductors more and more are applied in electric engineering. For their optimal applications it is necessary to be more familiar with their unique physical properties. In the present paper we will consider one of them it is peak effect in the dynamic current-voltage characteristics in slowly varying magnetic field. Previously peak effect was observed -in magnetization measurements. There are numerous methods of measurements conducted on superconducting materials, including especially high temperature oxide superconductors. The most frequent and popular methods are based on measurements of their resistive and magnetic properties. Magnetic characteristics of superconductors are related to very peculiar form of penetration of magnetic flux into them, that is in a form of vortices - flux lines or pancake shape. Elaborating new method for detection of magnetic induction penetration into HTc superconductors is therefore an important goal both from scientific reasons as well as possibility of construction of new superconducting electronic sensor. In the present paper analysis of the dynamic anomalies of current-voltage characteristics, arising as the result of the penetration of vortices in slowly varying magnetic field into HTc superconductors is discussed, also from the point of view of applications of these anomalies, as a new method for detecting properties of superconductors.

## II. MODELING DYNAMIC CURRENT-VOLTAGE CHARACTERISTICS ANOMALIES OF HTc SUPERCONDUCTORS IN A VARYING MAGNETIC FIELD

Current-voltage characteristics of superconducting materials allow to determine various material parameters of superconductors, related to their critical current. In a static magnetic field these current-voltage characteristics describe the decrease of the critical current with magnetic field, as shows Fig. 1 for $YBa_2Cu_3O_{7-x}$ sample, for which the superconducting state transition resistively measured is presented in Fig. 2. In slowly varying magnetic field there appear however in some cases anomalies, which have been observed us for instance in [1], while an example of their dependence on the transport current amplitude is shown in Fig. 3. These anomalies sensitive to magnetic field sweep rate, current amplitude, should be useful therefore as a possible tool for detecting magnetic quantities in superconductors – especially describing the magnetic flux penetration into HTc materials.

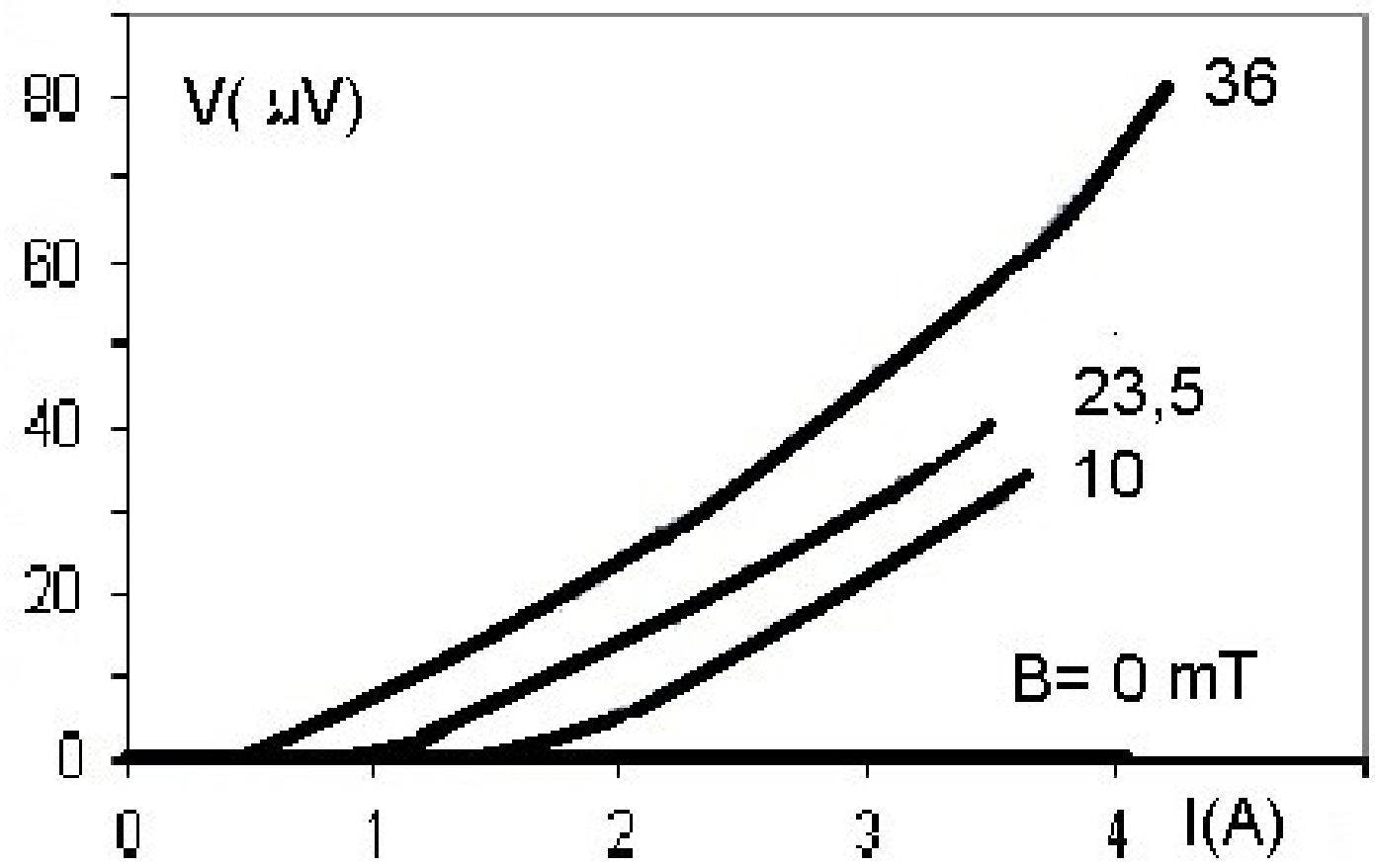

Fig. 1. Current-voltage characteristics of $YBa_2Cu_3O_{7-x}$ ceramic in external magnetic field at liquid nitrogen temperature

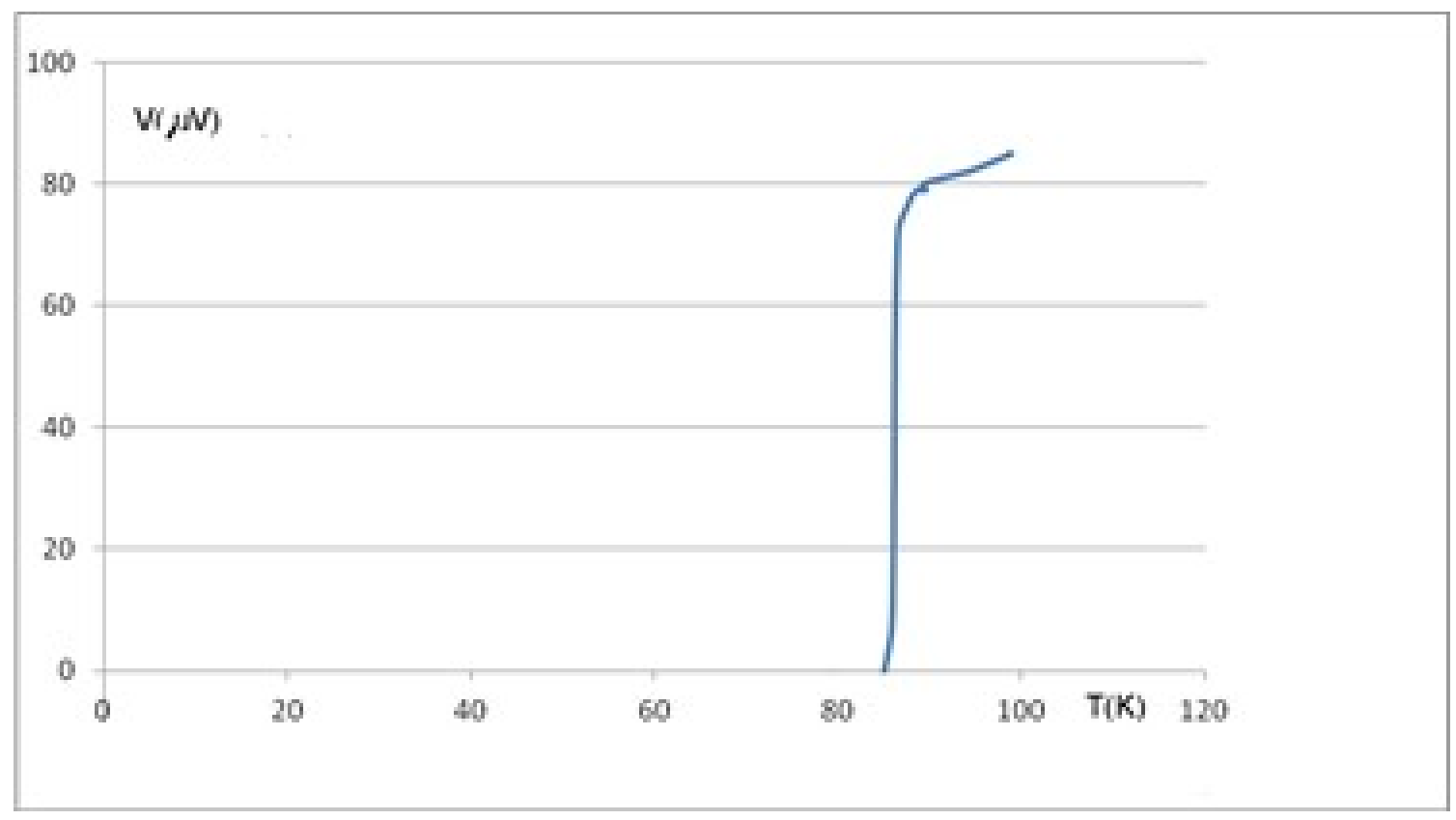

Fig. 2. Superconducting transition curve of YBaCuO sample

Experimentally above anomalies have been observed, while flat superconducting sample was placed into perpendicular to the surface, slowly increasing magnetic field. Then in some range of temperatures and magnetic field sweep rate, as well as transport current amplitudes the dynamic anomalies arise, as it was described in details previously in [1] and references here given. In the present paper a new theoretical analysis of these anomalies is given indicating their dependence on the diffusion of magnetic flux inside it.

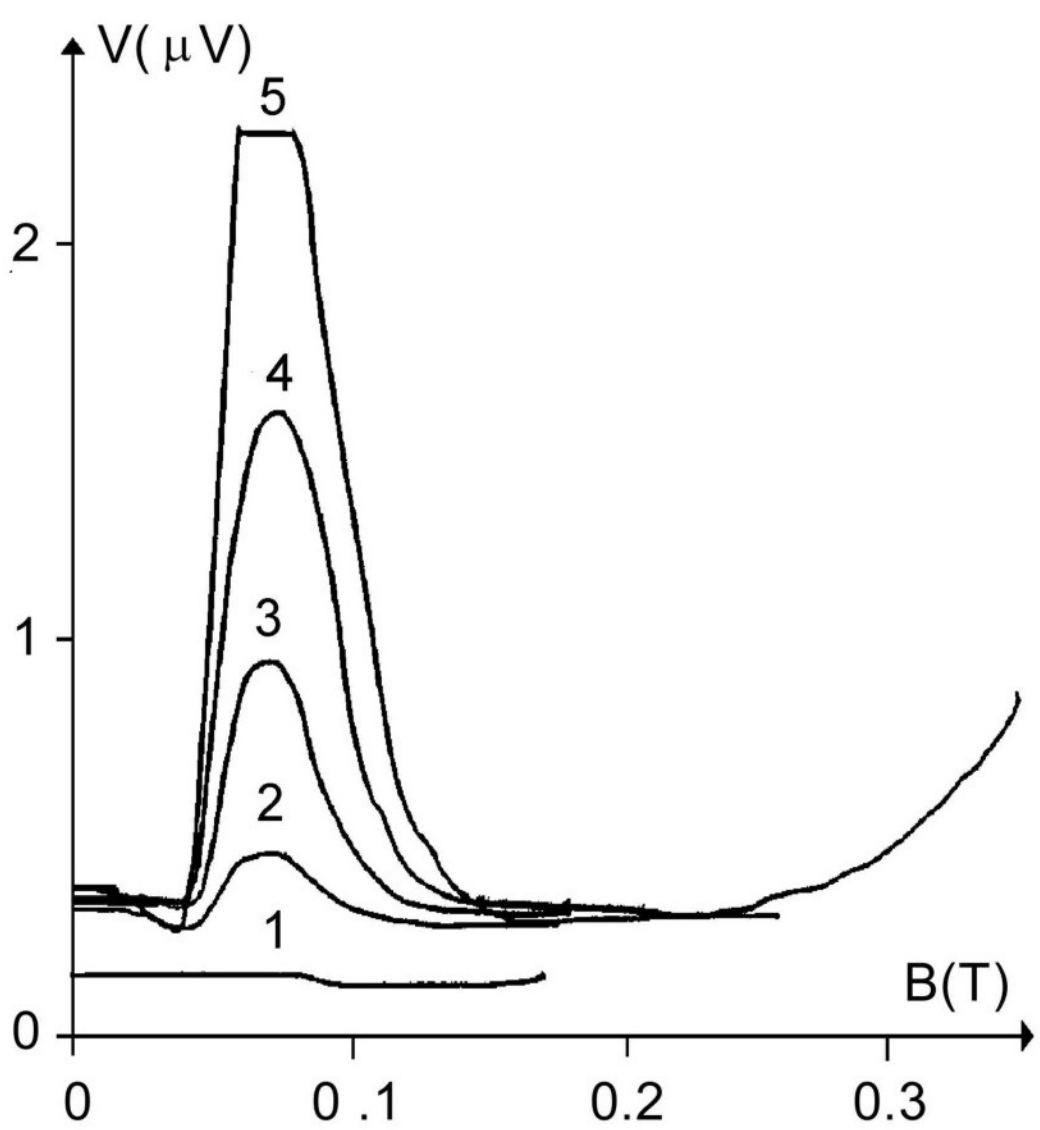

Fig. 3. Typical dynamic anomalies measured previously [1] of the I-V curves of YBaCuO superconductor in linearly slowly sweeping magnetic field for various values of transport current: (1) 110 mA, (2) 130 mA, (3) 140 mA, (4) 150 mA, (5) 160 mA and T= 45 K .

The present new approach to the interpretation of these anomalies is based on an analysis of the magnetic diffusion equation for the case of HTc superconducting plate inserted into linearly varying magnetic field perpendicular to the surface:

$$\frac{\partial B}{\partial t} = \frac{1}{D}\frac{\partial^2 B}{\partial x^2} \tag{1}$$

Equation 1 describes here the diffusion of the magnetic induction B(x,t) in the form of vortices into the superconductor with the diffusion coefficient D. In order to describe the experimental case of an applied linearly varying magnetic induction, perpendicular to the surface of superconductor, solution of diffusion Eq. 1 has been chosen in the following form:

$$B(x,t) = \frac{D\dot{B}}{2}x^2 + bx + B(0,t) \tag{2}$$

Symbol $\dot{B} = \dfrac{\partial B(0,t)}{\partial t}$ denotes time derivative of induction on the surface, which according to the experiment is constant, while parameter x is the depth inside the sample. Parameter b is related to the physical features of given process, which therefore determine its choice. Present case concerns superconductors, then critical current should decrease with magnetic field in each point of sample. The critical current is described now for that plate geometry according to Maxwell's equation, by the derivative $\partial B(x,t)/\partial x = D\dot{B}x + b$. In order to fulfill the condition of the decrease of the critical current with magnetic field also in this dynamic case function b has been chosen in the following form: $b = -\alpha D\dot{B}/B$. It leads then from physical reasons to necessary modification of the solution of the basic diffusion equation 2, dependent on the magnitude of the proportionality coefficient α. For small α we are approaching further tightly the exact solution of the diffusion equation. We should notice too, that in considered here case of the dynamic magnetic field the shielding current is not the same as static critical current. Magnetic induction increasing with time, penetrates the superconducting sample and reaches according to Eq. 2, its half-thickness $x_m$ at time $t_1$ given by the relation:

$$t_1 = \frac{-(0{,}5D\dot{B}x_m^2 + \Delta B) + \sqrt{\left(\frac{D\dot{B}x_m^2}{2} + \Delta B\right)^2 + 4D\dot{B}\alpha x_m}}{2\dot{B}} \tag{3}$$

Eq. 3 takes into account the shift of magnetic induction on the sample surface connected with flow of the transport current. Electric field E is then generated in the center of sample, where electric taps are put. Its sign depends on the respective orientation of the magnetic field and transport current density while the module is:

$$E = \frac{1}{(\dot{B}t)^2 t^2}\left[\dot{B}^3 x_m t^4 + \frac{\dot{B}^2 D\alpha x_m^2 t^2}{2} + \dot{B}\alpha\Delta B t^2 - D\alpha^3 + \left(\dot{B}^2 t^3 + D\alpha^2\right)\sqrt{\frac{D\alpha^2 - 2\dot{B}^2 t^3 - 2\dot{B}\Delta B t^2}{D}}\right] \tag{4}$$

$\Delta B$ appearing in Eqs. 3-4 is just mentioned above magnetic induction shift on the left side of the superconductor connected with transport current flow, while $-\Delta B$ respectively, is the value of this shift of induction on the opposite cover of the sample.

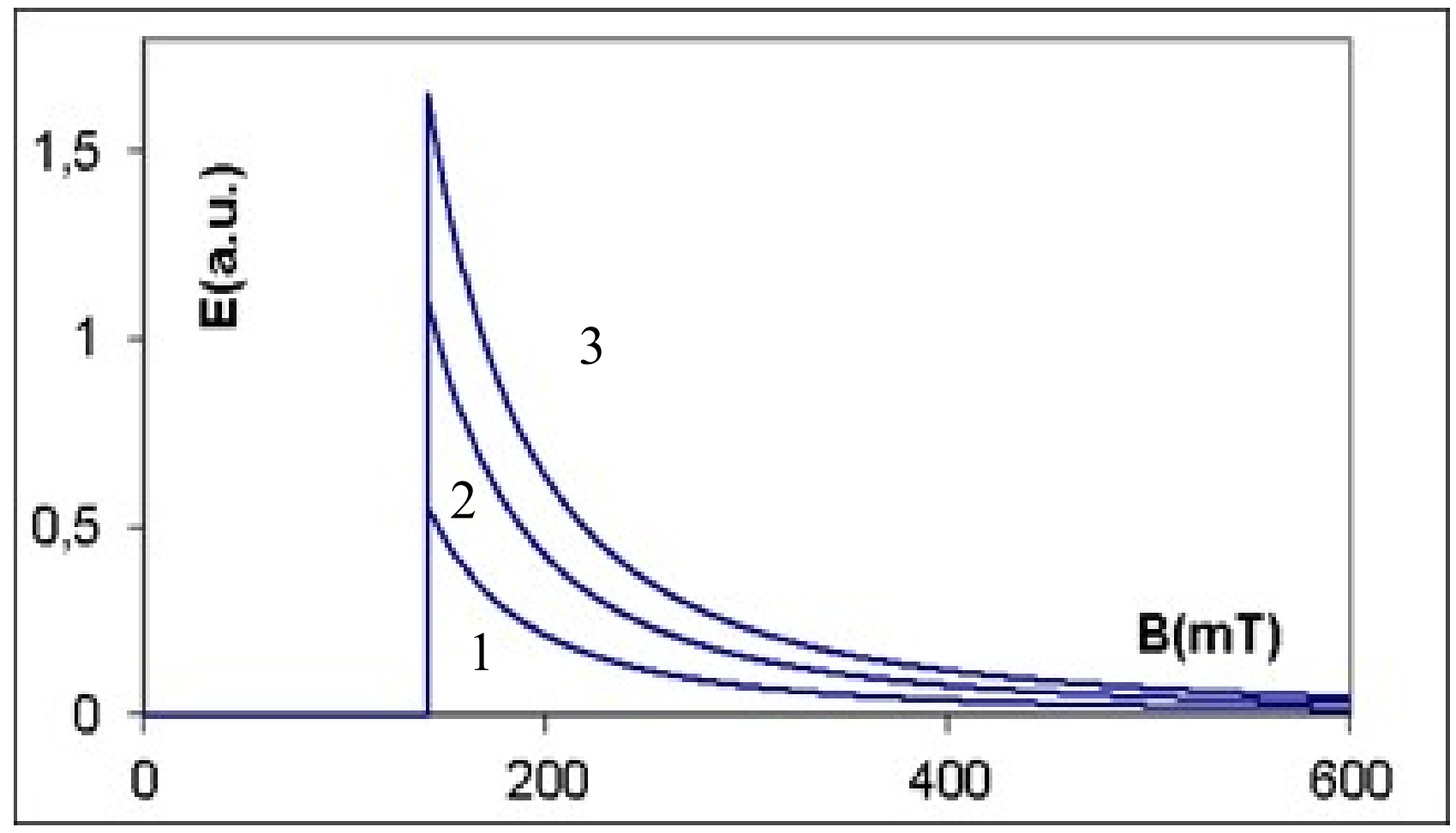

Fig. 4. Calculated dependence of the shape of the dynamical anomalies of I-V characteristics in varying magnetic field on the transport current amplitude: (1) 100 mA, (2) 200 mA, (3) 300 mA

Eq. 4 describes the non-saturated case, in which magnetic induction penetrates from one side of the superconductor up to the depth $x_m < x_1 < x_c$. $x_1$ is given here by the relation:

$$x_1 = \frac{\alpha}{B} - \frac{1}{\dot{B}}\sqrt{(\frac{\alpha}{t})^2 - \frac{2\dot{B}}{D}(B+\Delta B)} \tag{5}$$

while $x_c(t)$ is expressed by the formula:

$$x_c(t) = x_m + \frac{\Delta B}{D\left(\frac{\alpha}{t} - \dot{B}x_m\right)} \tag{6}$$

$x_c(t)$ is a function of time and describes the depth inside the sample at which both branches of magnetic induction coming from two edges of sample join together. This case is called here as the saturated one. The connection of two branches of magnetic induction leads to a change in the penetration of magnetic flux through the middle of the sample, where voltage taps are put and therefore to the variation of the generated electric field:

$$E = \frac{\alpha\Delta B}{2D^2 x_m\left(\dot{B}x_m t - \alpha\right)^4}\left\langle \begin{matrix} \dot{B}^3 D^2 x_m^5 t^2 - \dot{B}^2 D x_m^2 t(2D\alpha x_m^2 + t(2\alpha + \Delta B x_m)) + \\ \dot{B}x_m\left[D^2\alpha^2 x_m^2 + 2D\alpha t(2\alpha + \Delta B x_m) - (\Delta B t)^2\right] - D\alpha^2(2\alpha + \Delta B x_m) \end{matrix}\right\rangle + \frac{\Delta B}{D x_m} \tag{7}$$

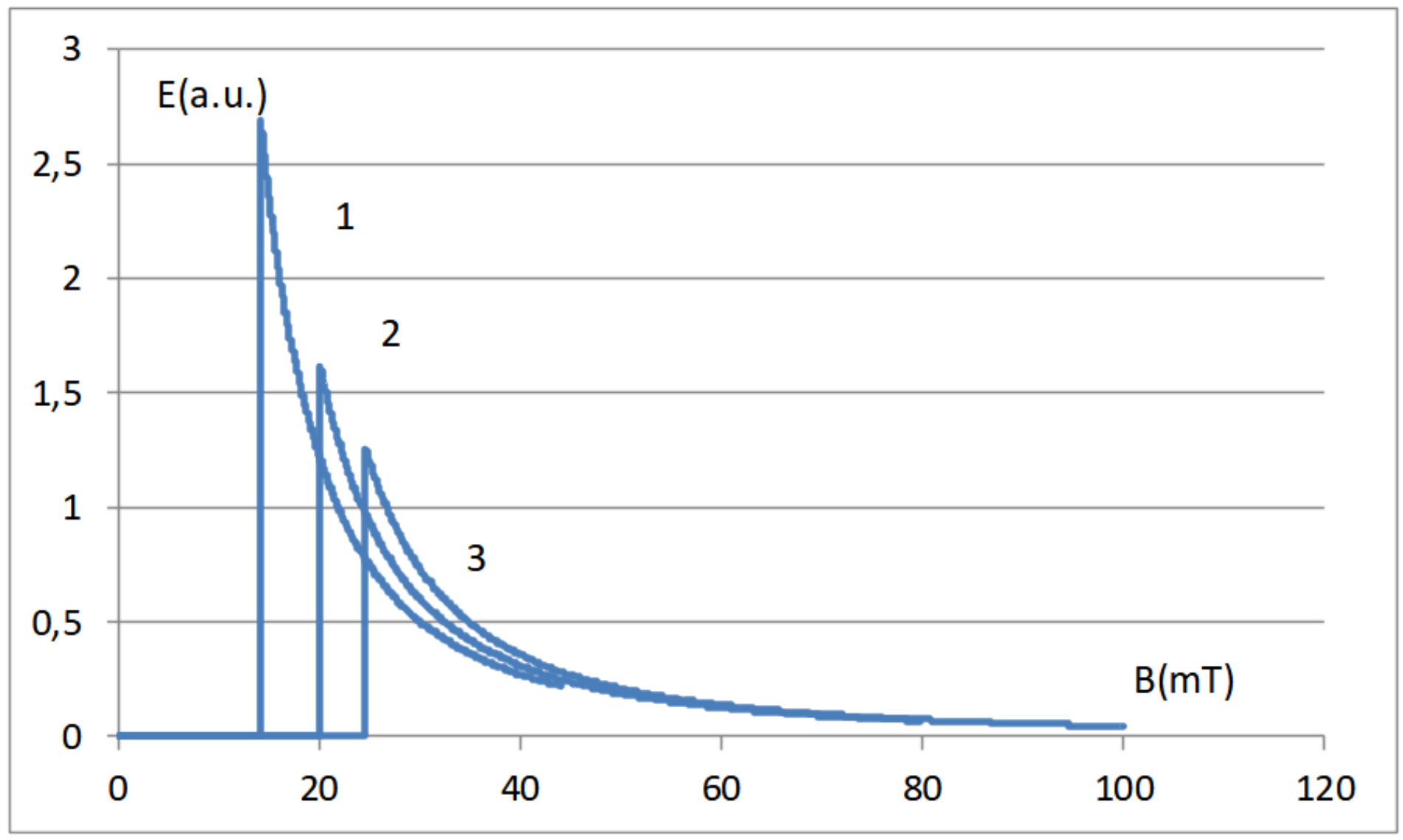

Fig. 5. Theoretically predicted influence of the magnetic field sweep rate on the dynamic anomalies of the current-voltage characteristics for HTc superconductors: (1) 10 mT/s, (2) 20 mT/s, (3) 30 mT/s

Time $t_2$ during which the transition from unsaturated to saturated state occurs is given by the condition:

$$D(x_m + \frac{\Delta B t_2}{D\alpha - x_m \dot{B} t_2})\left\langle \dot{B}(x_m + \frac{\Delta B t_2}{D\alpha - x_m \dot{B} t_2}) - \frac{2\alpha}{t_2} \right\rangle + 2\left( \dot{B} t_2 + \Delta B \right) = 0 \qquad (8)$$

In Fig. 4 is shown the result of the calculations of the influence of the transport current amplitude on the dynamic anomalies of the current-voltage characteristics in arbitrary units E(a.u), which indicates on an increase of magnitude of anomalies with transport current.

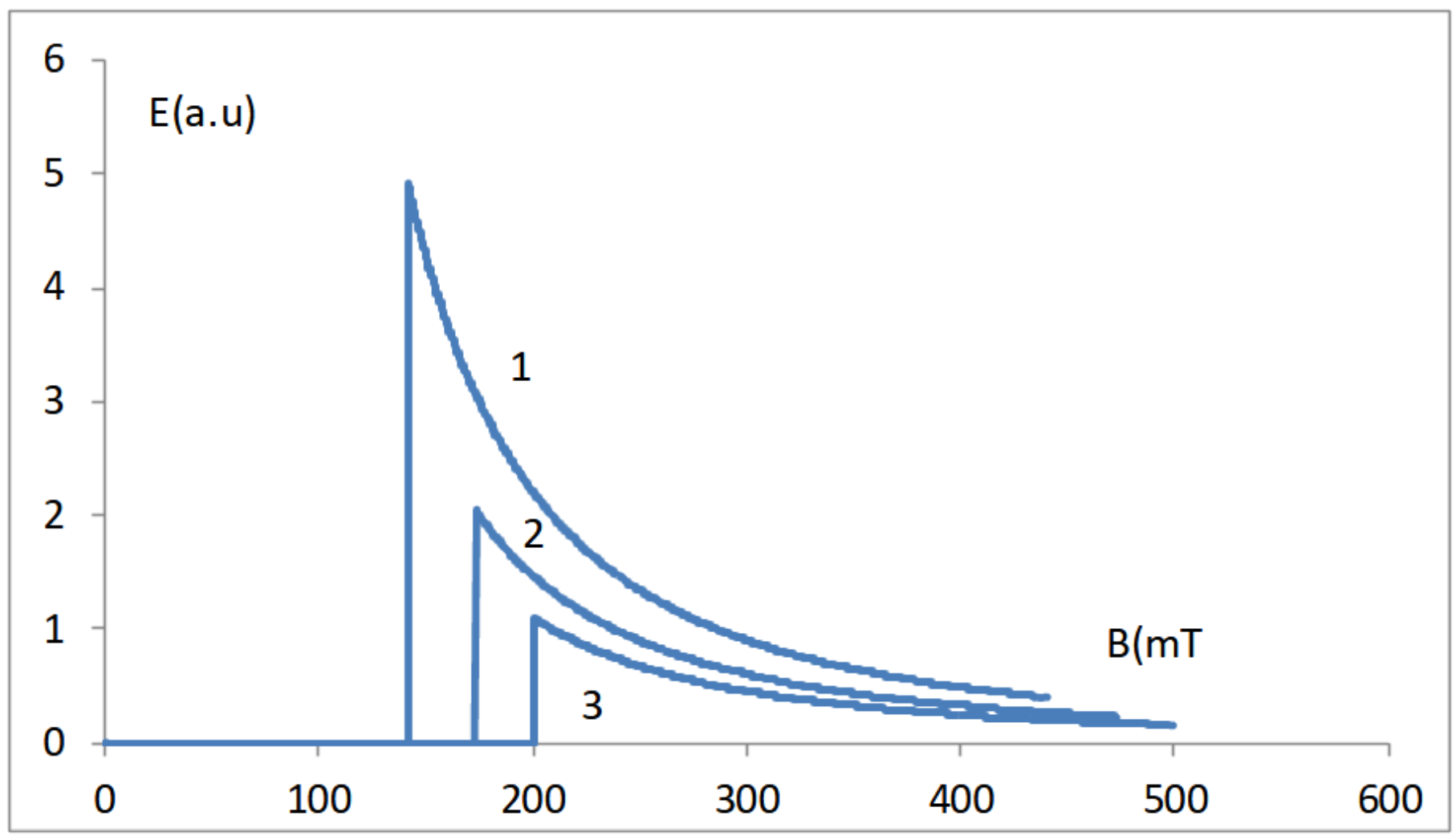

Fig. 6. Calculated influence of the magnetic diffusion coefficient on the dynamic anomalies of the current-voltage characteristics for HTc superconductors: (1) D=10, (2) 15, (3) 20 [s/m$^2$]

This dynamic effect of enhancement magnitude of anomalies with current has been really observed experimentally, as is presented in Fig. 3.

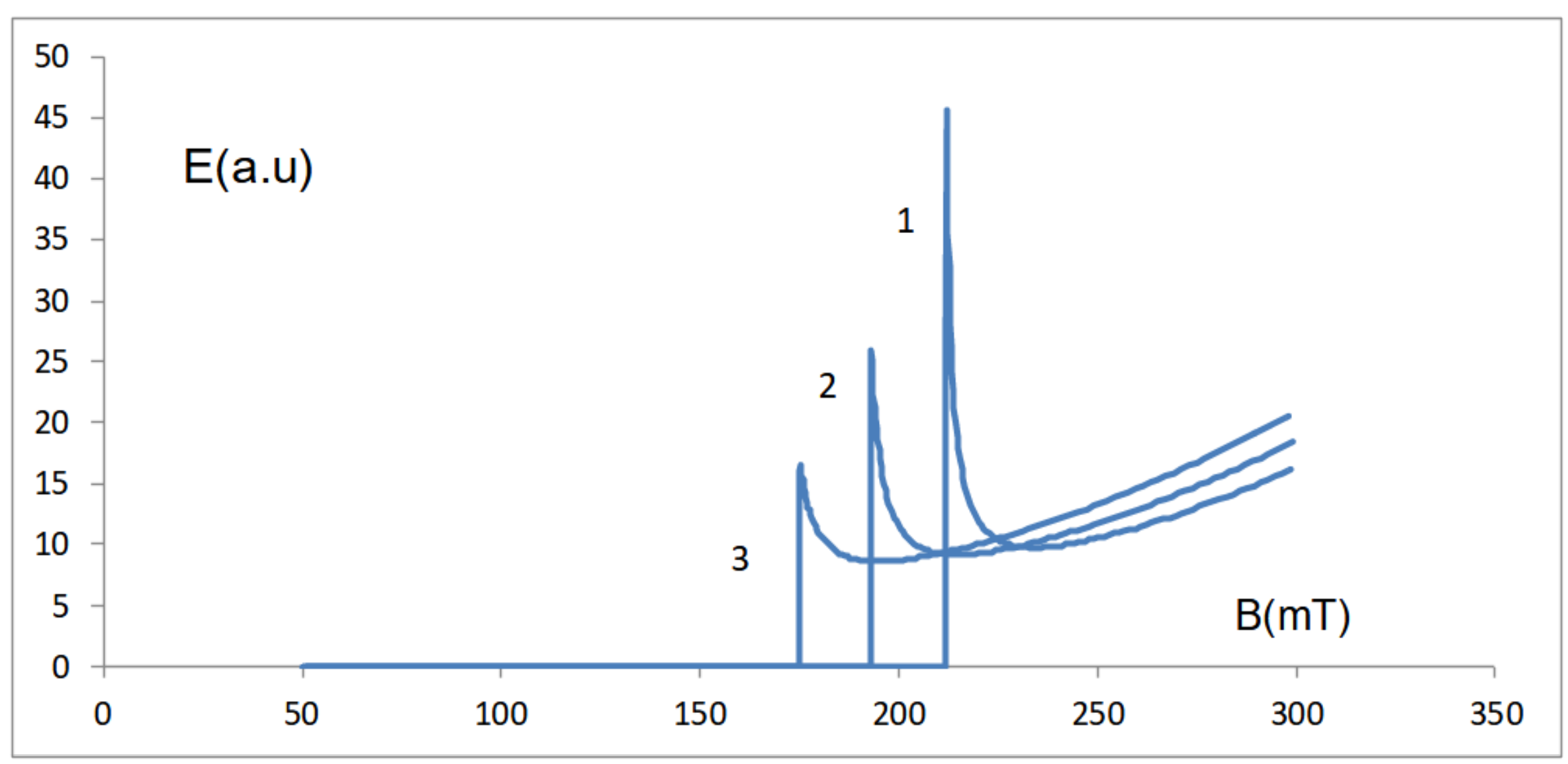

Fig. 7. The influence of superconducting material parameter $B^0$ : (1) $B^0$ = 0,12 T, (2) $B^0$ = 0,1 T, (3) $B^0$ = 0,08 T, appearing in Eq. 9 on dynamic anomalies of I-V curves in a slowly varying magnetic field, for current I = 5A Figure 5 from other side shows the influence, calculated according to this model, the change in the magnetic field sweep rate on the dynamic current-voltage characteristics anomalies. For a larger external magnetic field sweep rate the anomalies are shifted into a higher magnetic field, which is in accordance with experimental data [1]. In Fig. 6 is shown influence of diffusion coefficient D on dynamic current-voltage characteristics anomalies.

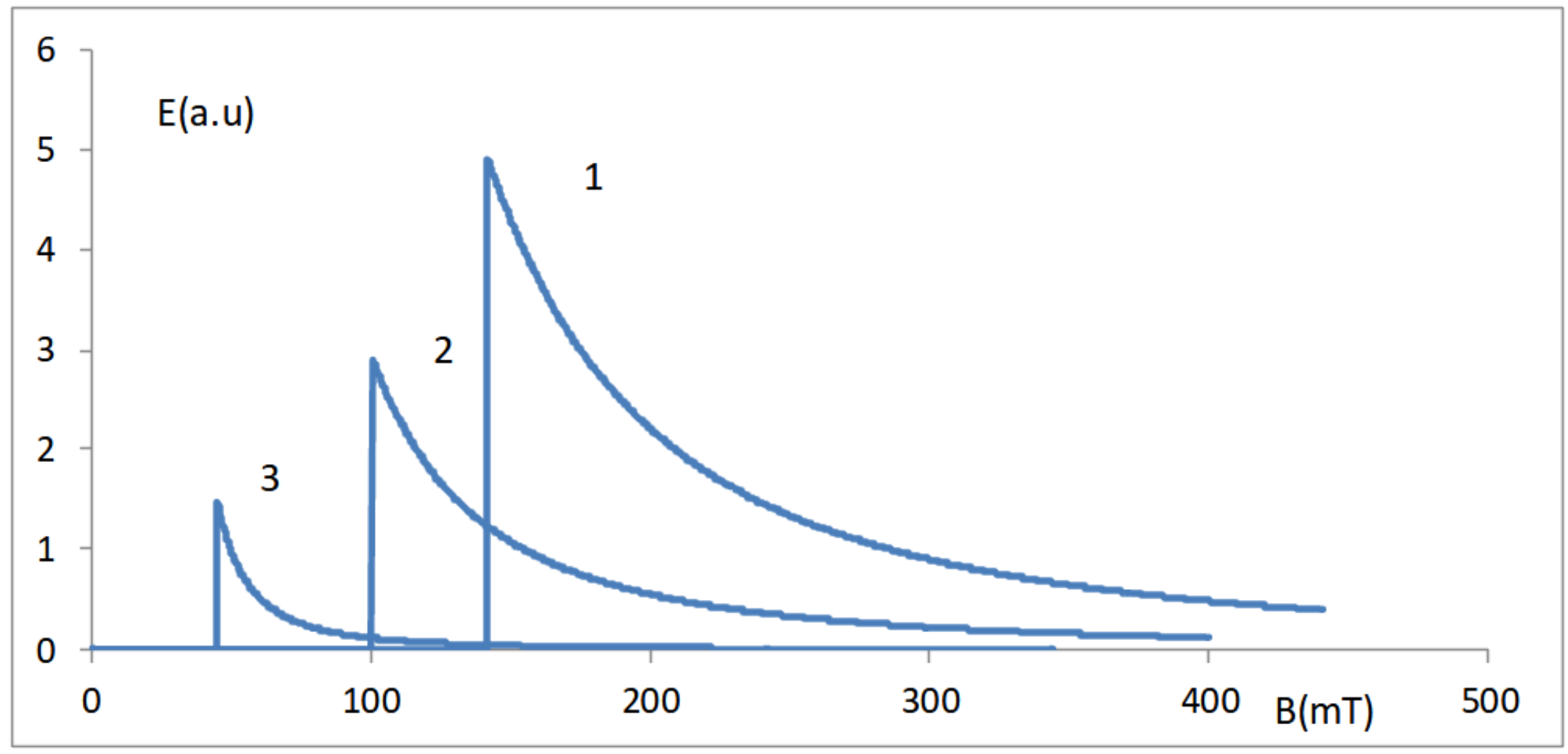

Fig. 8. The influence of material parameter β: (1) $\beta = 1\ T^{1+\gamma}/m$, (2) $\beta = 5$, (3) $\beta = 10$ on dynamic anomalies of I-V curves in a slowly varying magnetic field

In order to prove that this effect is related to the superconducting state properties it has been performed independent theoretical analysis based on critical state model, developed previously and presented for instance in [2]. The critical current magnetic field dependence is described then by generalized equation of the form:

$$\mu_0 j_c = \pm \frac{\beta}{\left(B(x) + B^0\right)^{\gamma}} \tag{9}$$

where $\mu_0$ is the magnetic permeability, β and $B^o$ are superconducting material parameters, while $\gamma = 1$ gives the Kim's critical state model [3] and $\gamma = 0$ describes the Bean's [4] approach.

As previously generated electric field has been calculated separately for non-saturated and saturated cases. In the non-saturated case, it is when both branches of the magnetic induction do not meet together and the superconductor is not penetrated fully by magnetic flux, the generated electric field is equal:

$$E = \frac{\dot{B}}{\beta}\left(B + \Delta B + B^0\right)^{\gamma} \cdot \left\langle \left[\left(B + \Delta B + B^0\right)^{1+\gamma} - \beta(1+\gamma)x_1\right]^{\frac{1}{1+\gamma}} - B^0 \right\rangle \tag{10}$$

For the magnetic induction totally penetrating HTc superconducting sample, the generated electric field is:

$$E = \frac{\dot{B}}{\beta}\left\langle \left(B + \Delta B + B^0\right)^{\gamma} \cdot \left( \left[ \left(B + \Delta B + B^0\right)^{1+\gamma} - \beta(1+\gamma)x_1 \right]^{\frac{1}{1+\gamma}} - B_{av}(x_m) \right) + B^{\otimes}\left(B_{av}(x_m) - B^0\right) \right\rangle \quad (11)$$

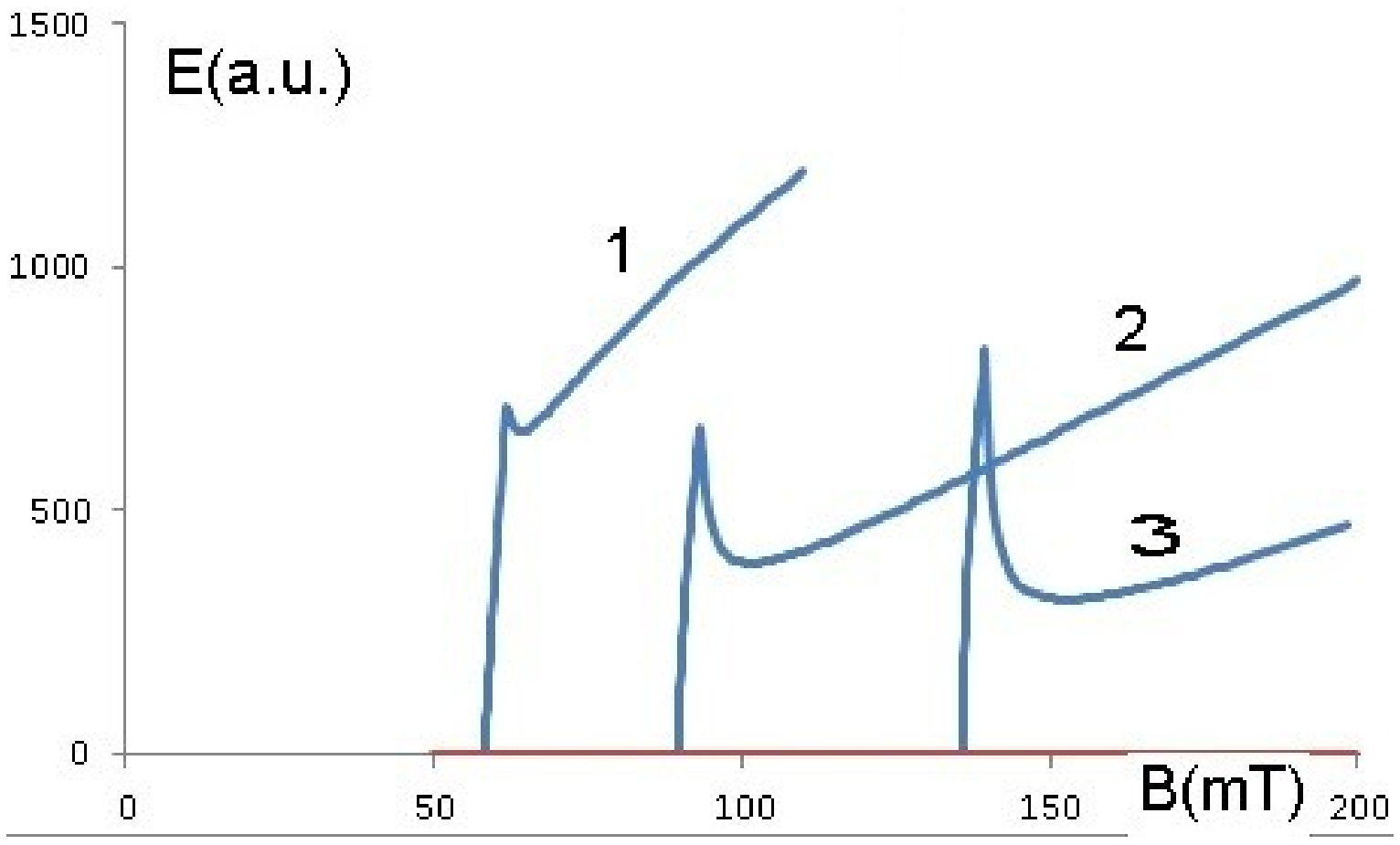

Fig. 9. The influence of material parameter, power exponent γ: (1) γ = 0.5, (2) γ = 1, (3) γ = 1,5 appearing in Eq. 9 on dynamic anomalies of I-V curves in a slowly varying magnetic field

In Eq. 11 magnetic quantities have been introduced:

$$B^{\otimes} = \frac{\left(B + \Delta B + B^0\right)^{\gamma} - \left(B - \Delta B + B^0\right)^{\gamma}}{2} \quad (12)$$

$$B_{av}(x_m) = \left[ \frac{\left(B + \Delta B + B^0\right)^{1+\gamma} + \left(B - \Delta B + B^0\right)^{1+\gamma}}{2} - \beta(1+\gamma)x_m \right]^{\frac{1}{1+\gamma}} \quad (13)$$

The results of calculations of the dynamic anomalies of the current-voltage characteristics according to the relations 9-13 are given in Figs. 7-9. This model explicitly takes into account the material parameters describing the HTc superconductor. Predicted theoretically dependence of the dynamic anomalies of the current-voltage characteristics on the electromagnetic parameters, such as transport current amplitude and material magnetic parameters $B^0$, γ, β describing magnetic field dependence of the critical current indicates on the general features of this phenomenon and therefore to the possibility to treat this effect as a tool for detecting magnetic quantities of the superconductors. While current-voltage characteristics allow to determine critical current of superconductors, this new method is sensitive as it follows from Fig. 3 to the dynamics of the magnetic flux penetration into these materials. Therefore this method should bring new information on the velocity of pancake vortices penetration and their diffusion, as well as generally stability behavior, which is an important parameter of superconducting materials and devices [5].

Acknowledgment
Author is grateful to Dr. V.I. Datskov for experimental collaboration.